\documentclass[12pt,preprint]{aastex}

\usepackage{rotating}

\shorttitle{A Novel Color Parameter for SNe Ia} \shortauthors{Wang
et al.}

\def\gsim{\;\lower4pt\hbox{${\buildrel\displaystyle >\over\sim}$}\;}
\def\lsim{\;\lower4pt\hbox{${\buildrel\displaystyle <\over\sim}$}\;}
\def\grls{\;\lower4pt\hbox{${\buildrel\displaystyle >\over <}$}\;}

\begin{document}

\title{A Novel Color Parameter As A Luminosity
Calibrator for Type Ia Supernovae}

\author{Xiaofeng Wang\altaffilmark{1,2},
Lifan Wang\altaffilmark{3}, Xu Zhou\altaffilmark{1}, Yu-Qing
Lou\altaffilmark{1,2,4}, Zongwei Li\altaffilmark{5}}
\altaffiltext{1}{National Astronomical Observatories of China,
Chinese Academy of Sciences, A20, Datun Road, Beijing 100012,
China} \altaffiltext{2}{Physics Department and Tsinghua Center for
Astrophysics (THCA), Tsinghua University, Beijing, 100084, China}
\altaffiltext{3}{Lawrence Berkeley National Laboratory 50-232, 1
Cyclotron Road, Berkeley, CA 94720, USA}
\altaffiltext{4}{Department of Astronomy and Astrophysics, the
University of Chicago, 5640 South Ellis Avenue, Chicago, IL 60637,
USA } \altaffiltext{5}{Astronomy Department, Beijing Normal
University, 19 Xinjiekouwai Dajie, Beijing, 100875, China}
\email{wxf@vega.bac.pku.edu.cn}

\begin{abstract}
Type Ia supernovae (SNe Ia) provide us with a unique tool for
measuring extragalactic distances and determining cosmological
parameters. As a result, the precise and effective calibration for
peak luminosities of SNe Ia becomes extremely crucial and thus is
critically scrutinized for cosmological explorations. In this
Letter, we reveal clear evidence for a tight linear correlation
between peak luminosities of SNe Ia and their $B-V$ colors $\sim
12$ days after the $B$ maximum denoted by $\Delta C_{12}$. By
introducing such a novel color parameter, $\Delta C_{12}$, this
empirical correlation allows us to uniformly standardize SNe Ia
with decline rates $\Delta m_{15}$ in the range of $0.8<\Delta
m_{15}<2.0$ and to reduce scatters in estimating their peak
luminosities from $\sim 0.5$ mag to the levels of 0.18 and 0.12
mag in the $V$ and $I$ bands, respectively. For a sample of SNe Ia
with insignificant reddenings of host galaxies [e.g.,
E$(B-V)_{host}\lsim 0.06$ mag], the scatter drops further to only
0.07 mag (or $3-4$\% in distance), which is comparable to
observational accuracies and is better than other calibrations for
SNe Ia. This would impact observational and theoretical studies of
SNe Ia and cosmological scales and parameters.
\end{abstract}


\section{Introduction}
The use of Type Ia supernovae (SNe Ia) as lighthouses or "standard
candles" for cosmological studies became feasible upon the
realization of an empirical correlation of peak luminosity with
light-curve shape. The discovery of the relationship between
light-curve shape and brightness led to a parameterization by the
decline rate $\Delta m_{15}$ in $B$-band brightness over 15 days
after the maximum light (Phillips 1993). The "multi-color light
curve shape" method (Riess et al. 1996, 1998) is based on
$\Delta$, the difference in peak luminosity between an observed SN
and a fiducial template. The "stretch" method (Perlmutter 1997;
Goldhaber et al. 2001) parameterizes light curves with a factor
$s$ which broadens or narrows the rest-frame timescale of a single
standard template to match with observed light curves. The
"Bayesian adapted template match" method introduced by Tonry et
al. (2003) estimates distances by comparing with a large set of
well-observed nearby SNe Ia rather than a parameterized template.
These methods are fundamentally equivalent by making use of
light-curve shapes. Another distance measurement technique is the
CMAGIC method (Wang et al. 2003), which uses the relationship
between light-curve shape and brightness in a more indirect way
than the above procedures. It is based on an observed linear
relationship in color-magnitude space for a certain period after
the maximum light to infer distances.

Besides the light-curve shape parameters, there are other
parameters that have been introduced to correlate with peak
luminosities of SNe Ia, such as line strengths of Ca and Si
absorption lines (Nugent et al. 1995), the Hubble types or colors
of parent galaxies (Hamuy et al. 1996; Branch et al. 1996), and
the galactocentric distance (Wang et al. 1997). These parameters
as peak luminosity descriptors are somewhat difficult to quantify,
especially for an SN Ia at a higher redshift $z$. In comparison,
the color $B-V$ at the maximum light seems to serve well as
another important peak luminosity descriptor, which has been
justified for further reducing statistical dispersions of distance
estimates when combined with decline rates (Tripp 1997; Tripp \&
Branch 1999; Parodi et al. 2000). As to those peculiar and
subluminous SNe Ia, the peak $B-V$ color was found to be a more
effective indicator of peak luminosity than the fading rate (e.g.,
Garnavich et al. 2004). The difficulty of using this color
parameter is that extremely accurate photometry and reddening
estimates are required, yet this situation may be mitigated for
the postmaximum $B-V$ color as differences in the color curves
become more conspicuous.

The main thrust of this Letter is to show that the $B-V$ colors of
SNe Ia, measured $\sim 12$ days after the $B$ maximum, strongly
correlate with their peak luminosities within a wide range of
decline rates. With this important empirical finding, it is now
possible to standardize all SNe Ia and use them to infer
extragalactic and cosmological distances with precisions
comparable to observational limits.

\section{The Novel Color Parameter $\Delta C_{12}$}

\subsection{The Color Curves}

Shown in Fig. 1 are the $B-V$ colors and their evolution of nine
well-sampled SNe Ia covering a broad range of observed light curve
decline rates. The $B-V$ color was $K$ corrected and de-reddened
by the Galactic reddening using the full-sky maps of dust infrared
emission (Schlegel et al. 1998). These SN Ia events most likely
suffer little or no reddening from dust in host galaxies, on the
basis of certain basic criteria such as the absence of
interstellar Na I or Ca II lines in the spectra, morphologies of
host galaxies, and the location of an SN in the host galaxy. In
reference to the light curves, the evolution of color curves
appears more complex and basically undergoes three stages: they
become bluer from the initial detection to about 5 days before the
maximum, then evolve redward by $\sim 1$ mag during $2-3$ weeks
after the maximum, and eventually recover from the reddest color
to zero when entering the nebular phase. Note that the $B-V$
colors of different SNe Ia converge $\sim 30$ days after the
maximum light and evolve in a similar way thereafter. This color
uniformity has proven very useful as an indicator of host galaxy
reddenings (Phillips et al. 1999; hereafter P99).

Except for the peculiar SNe Ia, the color curves of normal SNe Ia
appear to have similar shapes. Their $B-V$ colors approach zero
near the maximum light (Parodi et al. 2000). Nevertheless, a
careful examination of the early phase still reveals small shifts
among color curves of different SNe Ia. This may be attributed to
reddenings in host galaxies or to intrinsic differences among SNe
Ia (including the mass of accreting CO white dwarfs, C/O ratio,
metallicity, magnetic fields, etc.; e.g., Umeda et al. 1999;
Timmes et al. 2003; Lou 1994, 1995). The prominent feature of the
$B-V$ color evolution curve for those peculiar subluminous SNe
1991bg, 1998de, and 1999by is their unusually red $B-V$ colors in
the early phase. At the $B$ maximum, the $B-V$ colors for SNe
1991bg, 1998de, and 1999by are 0.69, 0.56, and 0.45 mag,
respectively. In addition to shifts among the $B-V$ color curves
of different SNe Ia, there are also variations in their shapes.
For example, SNe 1991bg and 1998de show dramatic declines and
reach their reddest colors earlier than normal SNe Ia do, at
$t\sim 12$ days after the $B$ maximum, rather than at $t\sim 30$
days for normal SNe Ia. Closer examinations of normal SNe Ia
reveal diverse decline rates in the color curves immediately after
the $B$ maximum (see Fig. 1). This diversity is apparently
associated with a slower flux variation in the $V$ band than in
the $B$ band for different SNe Ia during this phase.

\subsection{The Color Parameter $\Delta C_{12}$ as a Calibrator}

The $B-V$ color measured at the $B$ maximum is one of the most
important secondary parameters used to calibrate peak luminosities
of SNe Ia (Tripp 1997; Tripp \& Branch 1999; Parodi et al. 2000;
Garnavich et al. 2004) and even yields clues of host galaxy
reddenings together with the decline rate parameter $\Delta
m_{15}$ (P99). Note that peak values of $B-V$ mostly concentrate
in a range from $\sim -0.1$ to $\sim 0.1$ mag for normal SNe Ia.
Within such a narrow range, it is very difficult to precisely
measure the peak $B-V$ color. As apparent in Fig. 1, the $B-V$
colors $\sim 12$ days after the $B$ maximum clearly distinguish
the SN Ia color curves that vary from $\sim 0.1$ to $\sim 0.8$ mag
for normal SNe Ia and extend to $\sim 1.6$ mag for peculiar events
(SN 1998de) at this epoch. This postmaximum color not only
contains the intrinsic color shift at the maximum light but also
reflects, to a certain extent, the difference in light curves. We
emphasize the usage of the $B-V$ color 12 days after the $B$
maximum, referred to as $\Delta C_{12}$, as a powerful calibrator
for peak luminosities of SNe Ia.

The $\Delta C_{12}$ values for SNe Ia can be estimated directly
from the photometry or from the best-fitting template.
Uncertainties in $\Delta C_{12}$ include errors in observed
magnitudes, in foreground reddenings, and in the $K$ term. The
$\Delta C_{12}$ "colors" of 84 well-observed SNe Ia\footnote{The
photometric data are taken from Jha (2002 and references
therein).} are plotted against the decline rate parameter $\Delta
m_{15}$ in the left panel of Fig. 2. Those SNe Ia with light curve
coverage beginning $\gsim$7 days after the $B$ maximum were
excluded in our sample of SNe Ia, as their extrapolated maximum
magnitudes, and $\Delta C_{12}$ and $\Delta m_{15}$ parameters may
be inaccurate owing to larger errors.

\subsection{Host Galaxy Reddenings Derived from $\Delta C_{12}$}

Lira (1995) has shown that SNe Ia with low extinctions tend to
have the same $B-V$ colors between 30 and 90 days after the
maximum. This fact was exploited by P99 to estimate dust
reddenings of SNe Ia in host galaxies. Based on the late-time
colors, we find that 38 out of 84 SNe Ia have values of
E$(B-V)_{host}\lsim 0.06$ mag. We thus assume below that these SNe
Ia were essentially unreddened by dust in their host galaxies;
their $\Delta C_{12}$ values approximately represent the intrinsic
$B-V$ colors 12 days after the $B$ maximum.

As is clear in the left panel of Fig. 2, an empirical correlation
exists between the color parameter $\Delta C_{12}$ and the decline
rate $\Delta m_{15}$ for those SNe Ia with E$(B-V)_{host}\lsim
0.06$. This functional relation, covering the decline rates in the
range of $0.81\lsim\Delta m_{15}\lsim 1.95$, appears to be
monotonic and shows a rapid change or a "kink" in the slope near
$\Delta m_{15}\sim 1.65$. A cubic spline gives a reasonable fit to
the "low-reddening" points, with a dispersion of only 0.05 mag.
The fitting parameters are listed in Table 1. Despite this rapid
change in slope, the fast-declining events connect well with the
end of the `normal' color distribution. We utilize this $\Delta
C_{12}$ color$-\Delta m_{15}$ relation to infer dust reddenings of
SNe Ia in host galaxies (P99).

In the right panel of Fig. 2, we show the E$(B-V)_{host}$ values
inferred by our $\Delta C_{12}$ method as a function of host
galaxy reddenings derived by the P99 method for a sample of 84
well-observed SNe Ia. A very strong correlation exists between the
two reddening determinations, with no evidence for a difference in
the zero points (the weighted average of the difference amounts to
$0.005\pm0.041$). We are thus confident in using the $\Delta
C_{12}-\Delta m_{15}$ relation for estimating reddenings of SNe
Ia. The typical uncertainty in our color excess estimates is $\sim
0.07$ mag, slightly larger than that of P99; because of a stronger
dependence of $\Delta C_{12}$ on $\Delta m_{15}$, this uncertainty
may be as large as $\sim 0.2$ mag for SNe Ia with very high
$\Delta m_{15}$ values.

\section{The Color Versus Luminosity Relation}

The left half of Fig. 3 shows the absolute magnitudes of SNe Ia as
a function of the color parameter $\Delta C_{12}$. Both quantities
were corrected only for Galactic reddenings and $K$ terms. To
mitigate deviations from the Hubble expansion caused by peculiar
motions, we only considered a subsample of 63 SNe Ia with
redshifts $0.01\lsim z \lsim 0.1$. The distance to each SN Ia was
computed with redshifts of host galaxies (in the reference frame
of the cosmic microwave background) and with a Hubble constant
$H_{0}$=72 km s$^{-1}$ Mpc$^{-1}$ (e.g., Freedman et al. 2001;
Spergel et al. 2003). A possible peculiar velocity component of
600 km s$^{-1}$ was included in error bars of the absolute
magnitudes (e.g., Hamuy et al. 1996).

The resulting absolute magnitudes in the $B$, $V$, and $I$ bands
confirm the well-known fact that SNe Ia display a wide range of
peak luminosities. It is clear that SNe Ia with larger $\Delta
C_{12}$ (or redder color) are dimmer and vice versa. A remarkable
correlation emerges naturally as SN 1992bc and SN 1998de appear as
extreme source objects with blue and red colors, respectively.
This color$-$luminosity relation may be caused by combined effects
from reddenings in host galaxies and from intrinsic differences
among SNe Ia, as the slopes given by linear fits to the data are
only half of the values of the reddening coefficient $R$ observed
in the Milky Way. Dispersions around the best-fit lines in the
$B$, $V$, and $I$ bands are 0.25, 0.18, and 0.12 mag,
respectively. The most discrepant points correspond to the
severely reddened SNe 1995E, 1999gd and 2000ce, but this
discrepancy tends to disappear in the $I$ band where reddening
effects on luminosity become unimportant.

In the right panel of Fig. 3, this plot is repeated for a
subsample of 30 out of these 63 SNe Ia for which we find
insignificant reddenings from host galaxies [i.e.,
E$(B-V)_{host}\lsim 0.06$]. These diagrams reveal more clearly the
true nature of the peak luminosity$-$color parameter relations for
SNe Ia. Elimination of those SNe Ia with significant reddenings of
host galaxies not only decreases scatters in these correlations
but also shows these correlations to be linear in the $V$ and $I$
bands, and probably also in the $B$ band. These simple linear
correlations apply well to SNe Ia with different decline rates,
including the peculiar events SNe 1997cn, 1998bp, 1998de, and
1999da. Corresponding dispersions are 0.10, 0.07, and 0.07 mag for
the fits in the $B$, $V$, and $I$ bands, respectively. These
dispersions are comparable to observational errors and leave
little room for a dependence on other unknown parameters.

Table 2 contains the fitting parameters for the $\Delta
C_{12}-M_{max}$ linear correlations for the "low host galaxy
reddening" subsample of 30 SNe Ia. When the four subluminous SNe
Ia are excluded, the best-fit results remain nearly unchanged
except for the $B$ band, where the slope changes slightly to
1.74$\pm$0.19 and the dispersion decreases further down to 0.08
mag. The impressively low dispersion of these fits strongly
supports the validity of $\Delta C_{12}$ as an excellent peak
luminosity calibrator for SNe Ia. Moreover, this shows that normal
and peculiar SNe Ia may be calibrated in a unified manner and
argues for their likely common physical origin. Clearly, more
subluminous peculiar SNIa events are needed to firmly establish
this empirical correlation. The physics underlying this important
correlation remains to be explored and understood.

In contrast to $\Delta C_{12}$, the decline rate $\Delta m_{15}$
as a linear function of peak luminosity cannot account for a wider
range of peak luminosities of SNe Ia. A quadratic relation argued
by P99 may improve but still cannot fit well for those peculiar
SNe Ia with high $\Delta m_{15}$ values. We adopt an exponential
function (Garnavich et al. 2004) to match the decline rate with
peak luminosity for 30 SNe Ia in the right half of Fig. 3, which
leads to dispersions of 0.16, 0.12, and 0.10 mag for the fits in
the $B$, $V$, and $I$ bands, respectively. The much larger and
color-dependent dispersions indicate that a single $\Delta m_{15}$
parameter is inadequate to fit all SNe Ia, whereas the overall fit
can be significantly improved with a much reduced dispersion of
$\sim 0.10$ mag in each band by introducing an additional
parameter, namely, the $B-V$ color at the maximum light together
with the decline rate. This demonstrates again the crucial role
the color parameter plays in standardizing SNe Ia.

It might seem that the $\Delta C_{12}$ parameterization resembles
somewhat the CMAGIC method (Wang et al. 2003), as this color
parameter is close to the linear regime of the color magnitude
diagram (CMD). Both methods utilize the post-maximum color curves
to explore the color-magnitude relation of SNe Ia. However, the
former specifically emphasizes color differences at a given epoch,
while the latter tends to find out a more uniform magnitude at a
given color. In the CMAGIC approach, a magnitude is inferred from
the linear regimes of the CMD. In reference to peak magnitudes,
the CMAGIC magnitudes are more uniform because of a color
standardization, but they still depend strongly on $\Delta m_{15}$
(see fig. 10 of Wang et al. 2003); this means that they are still
influenced by light curve shapes. In contrast, this effect may
become the weakest around day 12 after the $B$ maximum when
maximum differences in color curves manifest among different SNe
Ia. In fact, this particular epoch corresponds to the time when
the brightest and the dimmest SNe Ia share the common CMAGIC
linear regime. For this reason, we strongly prefer to use the
$B-V$ color at this epoch, namely, $\Delta C_{12}$, as a key input
parameter for a more precise calibration of SNe Ia.

\section{Conclusions and Discussion}

We here demonstrate that $\Delta C_{12}$, the $B-V$ color measured
12 days after the $B$ maximum, provides effective discriminations
among color curves of SNe Ia. There is a stronger dependence of
$\Delta C_{12}$ on the decline rate $\Delta m_{15}$ of SNe Ia, and
this fact can be readily utilized for estimating reddenings of
host galaxies. The reddenings determined from the $\Delta
C_{12}-\Delta m_{15}$ relation agree remarkably well with those of
Phillips et al. (1999). This would be potentially important for
reddening estimates of those SNe Ia for which photometries are
sparse or unavailable at late times (e.g., 30 days $\lsim t\lsim$
90 days).

We have revealed a very tight linear correlation between $\Delta
C_{12}$ and peak luminosities of SNe Ia. A simple linear relation
seems to fit extremely well for both normal and peculiar SNe Ia,
with a small dispersion of $\lsim 0.10$ mag for a subsample of SNe
Ia, which suffer insignificant host galaxy reddenings. This much
reduced dispersion is almost comparable to observational error
limits. At this level, all observed SNe Ia can be more precisely
calibrated in terms of $\Delta C_{12}$. In comparison with $\Delta
m_{15}$, $\Delta C_{12}$ can be readily quantified as such, and
its relative error is smaller than that of $B-V$ at the maximum
light. More importantly, it incarnates the color variations near
the maximum light as well as the shape differences of the
postmaximum light curves. These empirical properties make $\Delta
C_{12}$ a more robust and reliable parameter for calibrating SNe
Ia, which will bear significant impact for using SNe Ia as
powerful standard candles for cosmological distances.

Given our success of using $\Delta C_{12}$ in calibrating peak
luminosities of broader SNe Ia with a reduced dispersion at low
redshifts $0.01\lsim z\lsim 0.1$, it would be natural to extend
this same color calibration for SNe Ia at higher redshifts $z\lsim
1.7$. While probable effects of interstellar extinction for SNe Ia
at $z\sim 1$ are not completely known, some preliminary empirical
results suggest that the extinction or reddening per hydrogen atom
decreases with increasing $z$ (e.g., York 2000). If this is
generally true except for peculiarities along particular lines of
sight, then our novel color $\Delta C_{12}$ calibration holds
great promise in determining the cosmological parameters more
accurately using the probe of SNe Ia in full power.

\acknowledgments Financial support for this work has been provided
by the National Science Foundation of China (NSFC grants 10303002
and 10173003) and the National Key Basic Research Science
Foundation (NKBRSF TG199075402). Y.Q.L. has been supported in part
by the ASCI Center for Astrophysical Thermonuclear Flashes at the
U.of Chicago under DoE contract B341495, by the Special Funds for
MSBSRP of China, by the Collaborative Research Fund from the NSFC
for Young Outstanding Overseas Chinese Scholars (NSFC 10028306) at
the NAOC, CAS, by NSFC grant 10373009 at the Tsinghua U., and by
the Yangtze Endowment from the MoE at the Tsinghua U. Affiliated
institutions of Y.Q.L. share this contribution.

\clearpage
\begin{figure}[htbp]
\figurenum{1}\hspace{0.3cm} \plotone{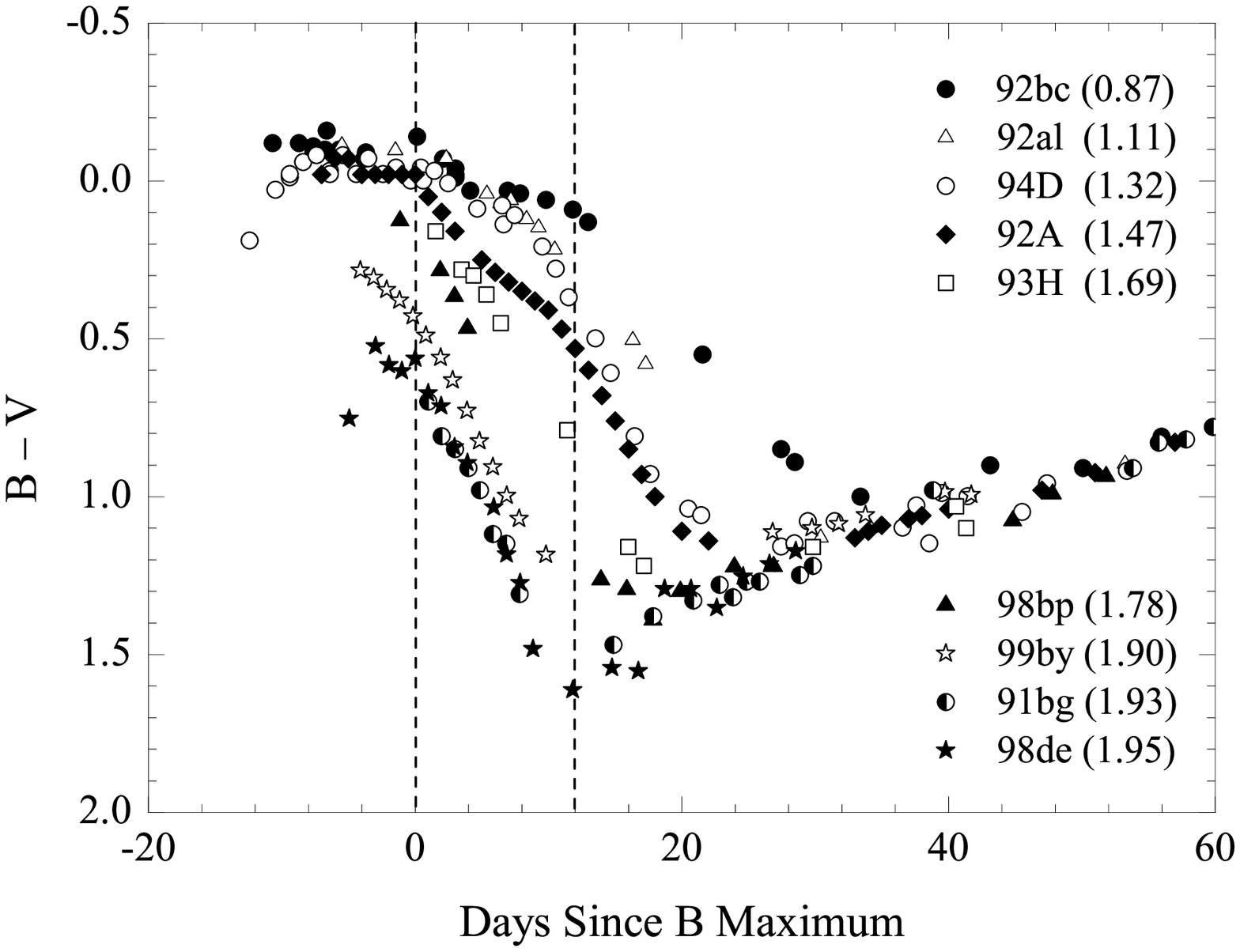}
\vspace{-0.2cm}
%\vspace{-0.3cm}
\caption{$B-V$ color evolution of nine SNe Ia most likely with
negligible reddening from dusts in host galaxies. These nine
events with their $\Delta m_{15}$ parameters indicated in
parentheses cover a wide range of initial decline rates and peak
luminosities. The two vertical dashed lines mark the epochs of the
$B$ maximum and 12 days after the $B$ maximum,
respectively.}\label{fig-1} \vspace{-0.0cm}
%\vspace{-0.5cm}
\end{figure}

\clearpage
\begin{figure}[htbp]
\figurenum{2} \plotone {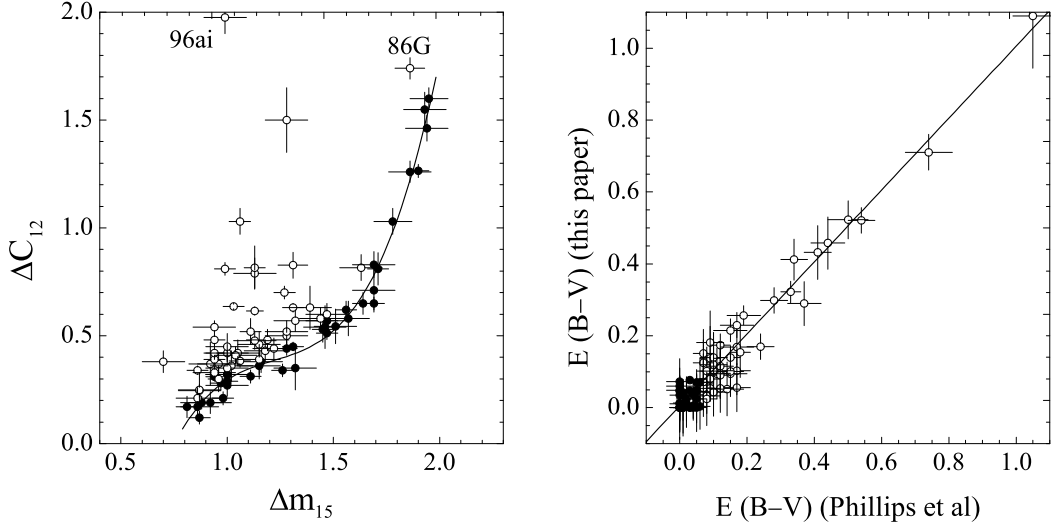} \vspace{-1.2cm} \hspace{-0.2cm}
\vspace{-1.5cm} \caption{{\it Left} $B-V$ colors $\Delta C_{12}$
of SNe Ia observed 12 days after the $B$ maximum plotted against
the decline rate parameter $\Delta m_{15}$. The open circles are
SNe Ia with E$(B-V)_{host}>0.06$ mag. The filled circles represent
38 events with E$(B-V)_{host}\lsim 0.06$ mag; these SNe Ia most
likely show little or no dust reddenings of host galaxies. The
solid curve is a cubic spline fit. {\it Right} Host galaxy
reddening derived from the $\Delta C_{12}-\Delta m_{15}$ relation
plotted against the estimates by P99.} \label{fig-2}
\vspace{-0.2cm}
\end{figure}

\clearpage
\begin{figure}
\figurenum{3} \hspace{-0.5cm} {\plotone{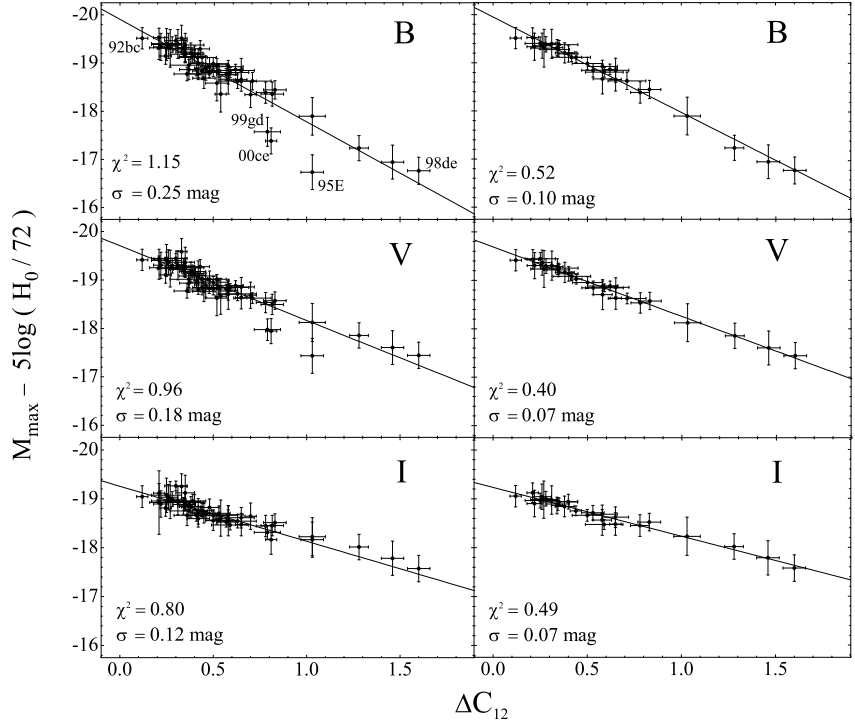}} \hspace{-0.5cm}
\caption{{\it Left} Absolute $B$, $V$, and $I$ magnitudes plotted
against the observed $\Delta C_{12}$ for 63 SNe Ia with redshifts
$0.01\lsim z\lsim 0.1$. Both quantities are corrected for Galactic
reddenings only. {\it Right} Same as the left panel, but
eliminating half of the sample with significant host galaxy
reddenings. The ridge lines are weighed fits to the points in each
panel.} \label{fig-3}
\end{figure}

\clearpage
\begin{table}
\caption{A cubic-spline fit to the $\Delta C_{12}-\Delta m_{15}$
relation $\Delta C_{12}=a+b_{1}(\Delta m_{15}-1.1)+b_{2}(\Delta
m_{15}-1.1)^{2} +b_{3}(\Delta m_{15}-1.1)^{3}$}
\begin{center}
\begin{tabular}{ccccc}
\tableline \tableline
a &$b^{a}_{1}$&$b^{a}_{2}$&$b^{a}_{3}$&$\sigma$(mag)\\
\tableline
0.347(016)&0.401(053)&$-$0.875(263)&2.440(280)&0.053\\
\tableline
\end{tabular}
\tablenotetext{a}{Error estimates in parentheses are in units of
$\pm 0^{m}.001$.}
\end{center}
\end{table}

\begin{table}
\caption{Fits to Color Parameter versus Luminosity Relation.}
\begin{center}
\begin{tabular}{ccccc}
\tableline \tableline
&&$M_{max}=M_{0}+R\Delta C_{12}$&\\
\cline{2-5}
Bandpass& $M^{a}_{0}$ &$R^{a}$ &$\sigma$ (mag)&n\\
\tableline
$B$ &$-19.96(07)$&1.94(13)&0.099&30\\
$V$ &$-19.72(07)$&1.46(12)&0.070&30\\
$I$ &$-19.23(07)$&1.03(13)&0.072&27\\
\tableline
\end{tabular}
\tablenotetext{a}{Error estimates in parentheses are in units of
$\pm 0^{m}.01$.} \tablenotetext{}{$n$ is the sample number of SNe
Ia selected.}
\end{center}
\end{table}
\end{document}